%% file: unfold.tex
\documentclass[12pt,a4paper]{article}
\usepackage[latin2]{inputenc}
\usepackage{t1enc}
\usepackage{amscd}
\usepackage{amsmath}
\usepackage{amssymb}
\usepackage{amsxtra}
\usepackage{setspace}
\usepackage{tabularx}
\usepackage{graphics}
\usepackage{graphicx}
\usepackage{fancyhdr}
\usepackage[thmmarks]{ntheorem-hyper}
\usepackage{blopeps}
\blopps{1mm}
\bloplw{0.1mm}

\theoremheaderfont{\bf}\theorembodyfont{\it}\theoremsymbol{\fbox{}}\theoremstyle{plain}\newtheorem{Thm}{Theorem}
\theoremheaderfont{\bf}\theorembodyfont{\rm}\theoremsymbol{\setlength{\unitlength}{1mm}\begin{picture}(2.5,2.5)\linethickness{2.5mm}\put(0,1.25){\line(1,0){2.5}}\end{picture}}\theoremstyle{plain}\newtheorem{Prf}{Proof}
\theoremheaderfont{\bf}\theorembodyfont{\it}\theoremsymbol{}\theoremstyle{plain}\newtheorem{Def}[Thm]{Definition}
\theoremheaderfont{\bf}\theorembodyfont{\sl}\theoremsymbol{}\theoremstyle{plain}\newtheorem{Rem}[Thm]{Remark}
\theoremheaderfont{\bf}\theorembodyfont{\it}\theoremsymbol{\fbox{}}\theoremstyle{plain}
\theoremheaderfont{\bf}\theorembodyfont{\sl}\theoremsymbol{}\theoremstyle{plain}
\theoremheaderfont{\bf}\theorembodyfont{\it}\theoremsymbol{\fbox{}}\theoremstyle{plain}

\begin{document}

\pagestyle{empty}

\input{unfold_00}

\pagestyle{fancy}

\setcounter{page}{1}

\lhead{}\chead{\scriptsize{A Robust Iterative Unfolding Method for Signal Processing}}\rhead{}

\lfoot{}\cfoot{\thepage}\rfoot{}

\input{unfold_01}

\input{unfold_02}

\input{unfold_03}

\input{unfold_04}

\input{unfold_05_ref}

\end{document}

%% file: unfold_00.tex
\begin{center}

\vspace*{5mm}

{\begin{spacing}{1}\LARGE\textbf{A Robust Iterative Unfolding Method for Signal Processing}\end{spacing}}

\vspace*{5mm}

{\large András LÁSZLÓ}

{MTA-KFKI Research Institute for Particle and Nuclear Physics, Budapest, Hungary}

{\texttt{laszloa@szofi.elte.hu}}

\vspace*{5mm}


\begin{abstract}
\noindent
It is a common problem in signal processing to remove a non-ideal detectors 
resolution from a measured probability density function of some physical quantity. 
This process is called unfolding (a special case is the deconvolution), and it 
would involve the inversion of the integral operator describing the folding (i.e.\ the 
smearing of the detector). 
Currently, there is no unbiased method known in literature for this issue (here, 
by unbiased we mean those approaches, which do not assume an ansatz for the 
unknown probability density function).

There is a well-known series expansion (Neumann series) in functional 
analysis for perturbative inversion of specific operators on Banach 
spaces. However, operators that appear in signal processing (e.g.\ 
folding and convolution of probability density functions), in general, 
do not satisfy the usual convergence condition of that series expansion. 
This article provides some theorems on the convergence criteria of 
a similar series expansion for this more general case, which is not 
covered yet by the literature.

The main result is that a series expansion provides a robust unbiased 
unfolding and deconvolution method. For the case of the deconvolution, 
such a series expansion can always be applied, 
and the method always recovers the maximum possible information about 
the initial probability density function, thus the method is optimal in 
this sense. A very significant advantage of the presented method is 
that one does not have to introduce \emph{ad hoc} frequency regulations etc., 
as in the case of usual naive deconvolution methods. For the case of 
general unfolding problems, we present a computer-testable sufficient 
condition for the convergence of the series expansion in question.

Some test examples and physics applications are also given. The most important 
physics example shall be (which originally motivated our survey on this topic) 
the case of $\pi^0\rightarrow\gamma+\gamma$ particle decay: we show that 
one can recover the initial $\pi^0$ momentum density function form the 
measured single $\gamma$ momentum density function by our series expansion.
\end{abstract}

\end{center}

%% file: unfold_01.tex
\section{Introduction}

In experimental physics, one commonly faces the following problem. 
The probability density function of a given physical quantity is to be 
measured (e.g.\ by histograming) with an experimental apparatus, but a 
non-ideal detector smears the signal. 
The question arises: if one knows the behavior of the detector quite well 
(i.e.\ one knows the response function of the detector), how can one 
reconstruct the original 
undistorted probability density function of the given physical quantity. 
Specially: 
there is an unknown probability density function $x\mapsto f(x)$ 
(this is the unknown probability density function of 
the undistorted 
physical quantity), and the measured density function is obtained by 
$y\mapsto g(y)=\int\rho(y|x)f(x)\;\mathrm{d}x$ (where the 
conditional density function $(y,x)\mapsto\rho(y|x)$ describes the 
smearing of the measurement apparatus, also called as 
\emph{response function}), then under which conditions and how can 
one re-obtain (i.e.\ \emph{unfold}) the original probability density function $f$ 
by measuring $g$ and by knowing $\rho$. We formalize this problem below. 
(In the text we shall abbreviate \emph{probability density function} by \emph{pdf}, 
\emph{conditional probability density function} by \emph{cpdf}, and the notion 
\emph{Lebesgue almost everywhere} or \emph{Lebesgue almost every}, known in measure 
theory, by \emph{ae}.)

Let $X$ and $Y$ be two finite dimensional real vector spaces, each equipped 
with the Lebesgue measure (which is unique up to a global positive constant 
factor). Then $L^{1}(X)$ and $L^{1}(Y)$ denote the space of Lebesgue integrable 
function classes $X\rightarrow\mathbb{C}$ and $Y\rightarrow\mathbb{C}$,
respectively.

\begin{Def} Let $\rho:Y\times X\rightarrow\mathbb{R}_{0}^{+},(y,x)
\mapsto\rho(y|x)$ is a cpdf over the product space $Y\times X$, 
(i.e.\ it is a nonnegative valued Lebesgue measurable function on the 
product space which satisfies for all 
$x\in X\,:\;\int\rho(y|x)\;\mathrm{d}y=1$). 
Then the linear operator 
\[A_{\rho}:\;L^{1}(X)\rightarrow L^{1}(Y),\quad
(x\mapsto f(x))\mapsto\left(y\mapsto\int\rho(y|x)f(x)\;\mathrm{d}x\right),\]
is called the \underline{folding operator by $\rho$}. 
\end{Def}

\begin{Rem}
The remarks below are trivial.
\begin{enumerate}
\item By Fubini's theorem, this linear operator is well defined.
\item By the monotonicity of integration, such an operator is continuous: 
\[\left\Vert A_{\rho}f \right\Vert_{L^{1}(Y)}=\int\left\vert\int\rho(y|x)f(x)\;\mathrm{d}x\right\vert\;\mathrm{d}y
\leq \int\int\rho(y|x)\left\vert f(x)\right\vert\;\mathrm{d}x\;\mathrm{d}y
=\left\Vert f \right\Vert_{L^{1}(X)}.\]
It is also trivial that we can saturate the above inequality by taking ae 
nonnegative function $f$, thus $\Vert A_{\rho}\Vert=1$ also follows.
\end{enumerate}
\end{Rem}

Our main interest will be the question: when is the operator $A_{\rho}$ 
invertible, and how the inverse operator could be evaluated on given pdfs 
in a constructive way.

\subsection{A special case: deconvolution problem}

A special case of the unfolding problem is the so called \emph{deconvolution}, 
i.e.\ when $Y=X$ and the cpdf $\rho$ is translation invariant in the sense that 
for all $a\in X$ and for all $y,x\in X\,:\;\rho(y|x+a)=\rho(y-a|x)$. In this case, 
the cpdf $\rho$ can be expressed by a pdf $\eta$ in the way 
$\rho(y|x)=\eta(y-x)$ for all $x,y\in X$. 

\begin{Def} Let $\eta$ be a pdf (i.e.\ it is a nonnegative valued Lebesgue integrable 
function on $X$ such that $\int\eta(x)\;\mathrm{d}x=1$). Then the linear operator 
\[A_{\eta}:\;L^{1}(X)\rightarrow L^{1}(X), \quad
f\mapsto \eta\star f =\left(y\mapsto\int\eta(y-x)f(x)\;\mathrm{d}x\right).\]
is called the  \underline{convolution operator by $\eta$}.
\end{Def}

We will state here a few properties of a convolution operator 
(see e.g.\ \cite{Arfken}, \cite{Bracewell}).

\begin{enumerate}
\item A convolution operator is not onto, and its image is not closed.
\item The range of a convolution operator 
is dense if and only if the Fourier transform of the convolver function 
is nowhere zero (\emph{Wiener's approximation theorem}).
\item A convolution operator is one-to-one if and only if the set of zeros 
of the Fourier transform of the convolver function has zero Lebesgue 
measure.
\end{enumerate}

\begin{Rem}
As a consequence, the inverse of a convolution operator -- if it exists at all --
is \emph{not} continuous. Indeed, the convolution operator is everywhere
defined and continuous, so it is closed, thus its inverse is
closed as well; since the domain of the inverse is not closed, the inverse
cannot be continuous by Banach's closed graph theorem.
\end{Rem}

We see that the characterization of a convolution operator is strongly related 
to the Fourier operators: 
\[F_{\pm}:\;L^1(X)\rightarrow C_{\infty}^{0}(X^{*}),\quad
\left(x\mapsto f(x)\right)\mapsto\left(y\mapsto\int
\mathrm{e}^{\pm\mathrm{i}\,\left<y|x\right>}f(x)\;\mathrm{d}x\right).\]
We denote by 
$C_{\infty}^{0}(X^{*})$ 
the space of continuous functions 
$X^{*}\rightarrow\mathbb{C}$ 
which have zero limit at the infinity. Here $X^{*}$ is the dual space of $X$, 
and for any $y\in X^{*}$ and $x\in X$ the number $\left<y|x\right>$ means the 
value of the covector $y$ on the vector $x$.

The Fourier operators have the following basic properties (\cite{Gasquet}):

\begin{enumerate}
\item $C_{\infty}^{0}(X^{*})$ is a Banach space with the maximum norm,
$F_{\pm}$ is continuous and $\Vert F_{\pm}\Vert=1$.
\item The Fourier operators are one-to-one. Thus,
the inverse Fourier operators $F_{\pm}^{-1}$ exist.
\item The range of $F_{\pm}$ is dense in $C_{\infty}^{0}(X^{*})$, 
however it is not the whole space. Thus, again by Banach's closed graph theorem, 
we infer that the operator $F_{\pm}^{-1}$ is \emph{not} continuous.
\item If $f,g\in L^{1}(X)$, then $F_{\pm}(f\star g)=
F_{\pm}(f)\cdot F_{\pm}(g)$ (\emph{convolution theorem}).
\end{enumerate}

The naive deconvolution procedure then goes in the following way:
\begin{enumerate}
\item take the Fourier transform of the convolution, 
$F_{\pm}(\eta\star f) $,
\item divide the above function by ${F_{\pm}\eta}$,
\item calculate the inverse Fourier transform;
\[f= F_{\pm}^{-1}\left(\frac{F_{\pm}(\eta\star f)}{F_{\pm}\eta}\right).\]
\end{enumerate}

The listed properties of the convolution operator, however, make it 
practically impossible to apply the deconvolution procedure 
in signal processing. The reason is that the measured density 
function (which is approximated by a normalized histogram in general) 
is not in the range of the convolution operator: 
it can be considered as the sum of a pdf in the range of the operator, 
plus a noise (e.g.\ Poissonian noise, originating from the statistical 
fluctuations of the entries in the histogram bins) outside the 
range of the operator in general. 
When applying the deconvolution procedure, the inverse operator can 
be calculated on the first term, however the deconvolution would give 
a nonsense result on the noise term, as it is 
not in the range of the convolution operator, thus leading to a 
nonsense result on the whole. 
Various noise suppression methods (high frequency cutoffs) are 
introduced as symptomatic treatment of this problem, however these 
solutions are based on rather intuitive approaches not on sound 
mathematics, and are highly non-unique (thus the derived solutions 
depend on the noise suppression approach). This is because the non-continuity 
of the inverse of the convolution operator: a small change caused by the 
high frequency regulation in the Fourier spectrum is not guaranteed to stay 
small after the deconvolution. 
This effect, in general, is referred to as: the deconvolution problem 
(or unfolding problem) is ill posed, i.e.\ one cannot get a 
robust method to do the deconvolution (or unfolding). Furthermore, 
if the Fourier transform of the convolver pdf has zeros in the finite, then 
the naive deconvolution becomes even more ambiguous: one has to introduce 
regulation procedures even at certain finite frequencies (at the 
zeros of the Fourier transform of the convolver pdf).

Despite of the above difficulties, we developed a robust perturbative 
method, which solves the problem. Our method of series 
expansion gives a robust and stable method for deconvolution. 
Using this method, the problem of zeros of the Fourier transform of the 
convolver pdf in the finite does not arise at all, 
furthermore one does not have to reconsider any high 
frequency regulations on a case-by-case intuitive basis. Plus, 
our series expansion is optimal in the sense that it recovers the 
maximum possible information about the initial pdf even in the case when the 
convolution in question is not even invertible.

%% file: unfold_02.tex
\section{Inverse operator by a series expansion}

There exists a basic theorem 
providing a perturbative method to obtain the inverse of continuous 
linear operators on a Banach space which are not too far from the 
identity operator. That theorem in its original form, however, 
does not apply to the case of convolution (or folding) operators. 
The main result of this paper is a generalization of that theorem 
to the case of convolution operators.

Now we recall the series expansion (called also Neumann series) for
the inverse of an operator.

Let  $A$ be a continuous linear operator on a Banach space such that 
$\Vert I-A \Vert< 1$, where $I$ is the identity operator. Then the 
operator $A$ is one-to-one and onto and its inverse is continuous, 
and the series 
$N\mapsto\sum_{n=0}^{N}(I-A)^{n}$ is absolutely  convergent to 
$A^{-1}$.

The proof is pretty simple, and can be found in any textbooks 
of functional analysis (e.g.\ \cite{Matolcsi}, \cite{Rudin}). It will be instructive, however, 
to cite the proof, as later we will strengthen this theorem.

First, it is easily shown by induction that $\sum_{n=0}^{N}(I-A)^{n}A=
A\sum_{n=0}^{N}(I-A)^{n}=I-(I-A)^{N+1}$.
The condition $\Vert I-A \Vert< 1$ guarantees that the sequence 
$N\mapsto(I-A)^{N+1}$ converges to zero in the operator norm, 
and the 
absolute convergence of the series $N\mapsto\sum_{n=0}^{N}(I-A)^{n}$, thus 
$\left(\sum_{n=0}^{\infty}(I-A)^{n}\right)A=
A\left(\sum_{n=0}^{\infty}(I-A)^{n}\right)=I$, i.e.\ $A^{-1}=\sum_{n=0}^{\infty}(I-A)^{n}$. 
As $A^{-1}$ is expressed as a limit of a series of continuous operators which 
is convergent in the operator norm, we infer that $A^{-1}$ is continuous.

\begin{Rem} The conditions of the above series expansion theorem 
fail for any folding operator $A_{\rho}$.
\begin{enumerate}
 \item We can observe that the series expansion is only meaningful for 
 the case of a folding operator only when the spaces $X$ and $Y$ are the same.
 \item Let us assume that $Y=X$. Then, it is easily obtained that a 
 folding operator $A_{\rho}$ does not  satisfy the required condition 
$\Vert I-A_{\rho} \Vert< 1$. It is 
trivial by the triangle inequality of norms that 
$\Vert I-A_{\rho} \Vert\leq 2$. We will show now that this inequality can be saturated for a wide class of cpdfs. 
Let us choose an arbitrary point $y\in X$, and consider the series of pdfs 
$n\mapsto\frac{1}{\lambda(K_{n}(y))}\chi_{{}_{K_{n}(y)}}$, 
where $K_{n}(y)$ are compact sets having non-zero 
Lebesgue measure $\lambda(K_{n}(y))$, such that $K_{n+1}(y)\subset K_{n}(y)$ for 
all $n\in \mathbb{N}$ and $\underset{n\in \mathbb{N}}{\cap}K_{n}(y)=\{y\}$.
Then, 
\[\left\Vert(I-A_{\rho})\frac{1}{\lambda(K_{n}(y))}\chi_{{}_{K_{n}(y)}}\right\Vert
=\int_{z\not\in K_{n}(y)}\int\rho(z|x)\frac{1}{\lambda(K_{n}(y))}\chi_{{}_{K_{n}(y)}}(x)\;\mathrm{d}x\;\mathrm{d}z\]
\[+\int_{z\in K_{n}(y)}\left\vert\frac{1}{\lambda(K_{n}(y))}\chi_{{}_{K_{n}(y)}}(z)-\int\rho(z|x)\frac{1}{\lambda(K_{n}(y))}\chi_{{}_{K_{n}(y)}}(x)\;\mathrm{d}x\right\vert\;\mathrm{d}z.\]
By making use of the fact that the integral of any pdf is $1$, one can write 
\[\int_{z\not\in K_{n}(y)}\int\rho(z|x)\frac{1}{\lambda(K_{n}(y))}\chi_{{}_{K_{n}(y)}}(x)\;\mathrm{d}x\;\mathrm{d}z=\]
\[1-\int\int\chi_{{}_{K_{n}(y)}}(z)\rho(z|x)\frac{1}{\lambda(K_{n}(y))}\chi_{{}_{K_{n}(y)}}(x)\;\mathrm{d}x\;\mathrm{d}z\]
for the first term. For the second term, one can use the monotonity of integration:
\[\int_{z\in K_{n}(y)}\left\vert\frac{1}{\lambda(K_{n}(y))}\chi_{{}_{K_{n}(y)}}(z)-\int\rho(z|x)\frac{1}{\lambda(K_{n}(y))}\chi_{{}_{K_{n}(y)}}(x)\;\mathrm{d}x\right\vert\;\mathrm{d}z\]
\[\geq \left\vert\int_{z\in K_{n}(y)}\left(\frac{1}{\lambda(K_{n}(y))}\chi_{{}_{K_{n}(y)}}(z)-\int\rho(z|x)\frac{1}{\lambda(K_{n}(y))}\chi_{{}_{K_{n}(y)}}(x)\;\mathrm{d}x\right)\;\mathrm{d}z\right\vert\]
\[=\left\vert\int\frac{1}{\lambda(K_{n}(y))}\chi_{{}_{K_{n}(y)}}(z)\;\mathrm{d}z-\int\int\chi_{{}_{K_{n}(y)}}(z)\rho(z|x)\frac{1}{\lambda(K_{n}(y))}\chi_{{}_{K_{n}(y)}}(x)\;\mathrm{d}x\;\mathrm{d}z\right\vert\]
\[=\left\vert 1-\int\int\chi_{{}_{K_{n}(y)}}(z)\rho(z|x)\frac{1}{\lambda(K_{n}(y))}\chi_{{}_{K_{n}(y)}}(x)\;\mathrm{d}x\;\mathrm{d}z\right\vert\]
\[=1-\int\int\chi_{{}_{K_{n}(y)}}(z)\rho(z|x)\frac{1}{\lambda(K_{n}(y))}\chi_{{}_{K_{n}(y)}}(x)\;\mathrm{d}x\;\mathrm{d}z\]
Here, at the second equality $\int\frac{1}{\lambda(K_{n}(y))}\chi_{{}_{K_{n}(y)}}(z)\;\mathrm{d}z=1$ was used, and 
the fact that the integral of any pdf over a Borel set is smaller or equal to $1$ 
was used at the third equality. Thus, we infer the inequality:
\[\left\Vert(I-A_{\rho})\frac{1}{\lambda(K_{n}(y))}\chi_{{}_{K_{n}(y)}}\right\Vert
\geq 2\cdot\left(1-\int\int\chi_{{}_{K_{n}(y)}}(z)\rho(z|x)\frac{1}{\lambda(K_{n}(y))}\chi_{{}_{K_{n}(y)}}(x)\;\mathrm{d}x\;\mathrm{d}z\right).\]
If the point $(y,y)\in X\times X$ is a Lebesgue point of $\rho$, then we will 
shown that the integral term goes to zero when $n$ goes to infinity, thus saturating 
our inequality in question. If a function $g:X\rightarrow\mathbb{C}$ is locally integrable, 
then a point $y\in X$ is called a \emph{Lebesgue point of $g$} if $\underset{n\rightarrow\infty}{\mathrm{lim}}\;\frac{1}{\lambda(K_{n}(y))}\int_{K_{n}(y)} \vert g(x)-g(y)\vert\;\mathrm{d}x=0$. If 
$y\in X$ is a Lebesgue point for $g$, then by the monotonity of integration it also follows that 
$\underset{n\rightarrow\infty}{\mathrm{lim}}\;\frac{1}{\lambda(K_{n}(y))}\int_{K_{n}(y)} g(x)\;\mathrm{d}x=g(y)$. 
Applying this result for $\rho$ on the product space $X\times X$ (assuming that the point $(y,y)\in X\times X$ is a Lebesgue point of $\rho$), 
we have that the sequence $n\mapsto\frac{1}{\lambda(K_{n}(y))}\frac{1}{\lambda(K_{n}(y))}\int_{K_{n}(y)}\int_{K_{n}(y)}\rho(z|x)\;\mathrm{d}x\;\mathrm{d}z$ is convergent to $\rho(y|y)$. 
Multiplying this sequence by the sequence $n\mapsto \lambda(K_{n}(y))$ (which is convergent to zero), we 
infer that $\underset{n\rightarrow\infty}{\mathrm{lim}}\;\frac{1}{\lambda(K_{n}(y))}\int_{K_{n}(y)}\int_{K_{n}(y)}\rho(z|x)\;\mathrm{d}x\;\mathrm{d}z=0$. 
If $\rho$ is continuous, then every point in $X\times X$ is a Lebesgue point of $\rho$. 
Thus, we have shown that if the cpdf $\rho$ is continuous, then $\Vert I-A_{\rho} \Vert=2$ holds, therefore 
the original theorem of Neumann cannot be applied directly for a folding operator with continuous cpdf.
\end{enumerate}
\end{Rem}

Apart from the above remark, the reason is obvious for the obstruction 
of inverting the convolution on the operator level: as the convolution
operators are not onto in general, one only can try to invert the operator 
on a function in the range of the operator. 
We try to modify the theorem for the case of convolution operators 
requiring, instead of convergence in the operator series, 
the convergence of the series  $N\mapsto\sum_{n=0}^{N}(I-A)^{n}(Af)$ in 
some sense (equivalently, the convergence of the sequence 
$N\mapsto(I-A)^{N+1}f$ in the same sense), for any $f\in L^{1}(X)$.

For getting a convenient result, let us recall that the elements of $L^{1}(X)$
can be viewed as regular tempered distributions. 
The Fourier transformations can be extended to the space of tempered 
distributions, where they are one-to-one and onto, continuous, and their 
inverse is also continuous (\cite{Matolcsi}, \cite{Rudin}). 
The proof of convergence will be performed on the Fourier transforms of the
functions, then the result will be brought back by using the continuity of 
the inverse Fourier transformation on the space of tempered distributions.

\begin{Thm}
Let $A_{\eta}$ be a convolution operator for some $\eta\in L^{1}(X)$. 
Let $Z$ be the set of zeros of the function $F_{\pm}\eta$. 
If the inequality 
\[\left\vert 1-F_{\pm}\eta \right\vert < 1\]
is satisfied everywhere outside $Z$, then for all $f\in L^{1}(X)$ the series 
\[N\mapsto\sum_{n=0}^{N}(I-A_{\eta})^{n}(A_{\eta}f)\]
is convergent in the space of tempered distributions, and 
\[\sum_{n=0}^{\infty}(I-A_{\eta})^{n}(A_{\eta}f)=f-F_{\pm}^{-1}(\chi_{{}_{Z}}F_{\pm}f).\]
\end{Thm}
\begin{Prf}
Assume that $\left\vert 1-F_{\pm}\eta \right\vert < 1$ holds everywhere outside 
$Z$. Let $V$ denote the subset of $X^{*}$ where $F_{\pm}\eta$ is nonzero. 
It is clear that $V$ and $Z$ are disjoint Lebesgue measurable sets and 
$X^{*}=V\cup Z$. 
Trivially, the sequence $N\mapsto \left\vert 1-F_{\pm}\eta \right\vert^{N+1}$ 
converges pointwise to $0$ on $V$, furthermore 
$\left\vert 1-F_{\pm}\eta \right\vert^{N+1}=1$ on $Z$ for all $N$. 
For every $f\in L^{1}(X)$ and rapidly decreasing test function $\varphi$ 
on $X^{*}$, we have 
\[\left\vert\int(1-F_{\pm}\eta(y))^{N+1}F_{\pm}f(y)
\cdot\varphi(y)\;\mathrm{d}y-\int\chi_{{}_{Z}}\cdot F_{\pm}f(y)
\cdot\varphi(y)\;\mathrm{d}y\right\vert=\]
\[\left\vert\int_{V}(1-F_{\pm}\eta(y))^{N+1}F_{\pm}f(y)
\cdot\varphi(y)\;\mathrm{d}y\right\vert\leq\]
\[\int_{V}\left\vert1-F_{\pm}\eta(y)\right\vert^{N+1}
\left\vert F_{\pm}f(y)\right\vert\cdot\left\vert\varphi(y)\right\vert\;\mathrm{d}y.\]
The series of Lebesgue integrable functions 
$N\mapsto\left\vert1-F_{\pm}\eta\right\vert^{N+1}
\left\vert F_{\pm}f\right\vert\cdot\left\vert\varphi\right\vert$ 
converges pointwise to zero on $V$, and 
$\left\vert1-F_{\pm}\eta\right\vert^{N+1}
\left\vert F_{\pm}f\right\vert\cdot\left\vert\varphi\right\vert\leq
\left\vert1-F_{\pm}\eta\right\vert^{1}
\left\vert F_{\pm}f\right\vert\cdot\left\vert\varphi\right\vert$ 
for all $N$, 
thus by Lebesgue's theorem of dominated convergence the last term of the 
inequality tends to zero when $N$ goes to infinity. Therefore, 
the function series $N\mapsto(1-F_{\pm}\eta)^{N+1}(F_{\pm}f)$ is convergent 
in the space of tempered distributions to the function 
$\chi_{{}_{Z}}F_{\pm}f$. Applying the inverse Fourier transformation 
$F_{\pm}^{-1}$ and using the 
continuity of the inverse Fourier transformation in the space of tempered 
distributions, we get the 
desired result, as by the convolution theorem we have 
$F_{\pm}^{-1}\left((1-F_{\pm}\eta)^{N+1}(F_{\pm}f)\right)=(I-A_{\eta})^{N+1}f$, and because 
\[f-\sum_{n=0}^{N}(I-A_{\eta})^{n}(A_{\eta}f)=(I-A_{\eta})^{N+1}f\]
for all $N$.
\end{Prf}
\begin{Rem} Let us assume that the condition of the theorem holds. Then it 
is quite evident that
\begin{enumerate}
\item If $Z$ has zero Lebesgue measure (which holds if and only if 
$A_{\eta}$ is one-to-one), then $F_{\pm}^{-1}(\chi_{{}_{Z}}F_{\pm}f)=0$. 
This means that the series in question always restores the arbitrarily chosen 
original function $f$ if and only 
if $A_{\eta}$ is one-to-one, i.e.\ if and only if $F_{\pm}\eta$ is ae 
nowhere zero.
\item If $Z$ has nonzero Lebesgue measure, our series also converges, 
and restores the maximum possible information about the original function 
$f$, namely the tempered distribution 
$f-F_{\pm}^{-1}(\chi_{{}_{Z}}F_{\pm}f)$. However, this tempered 
distribution may not be a function in general. If the function 
$\chi_{{}_{Z}}F_{\pm}f$ is not a continuous function which tends to zero at 
the infinity, then 
$F_{\pm}^{-1}(\chi_{{}_{Z}}F_{\pm}f)$ cannot be an integrable function. 
As we shall see in the next section, if the function 
$\chi_{{}_{Z}}F_{\pm}f$ is not a continuous function which is bounded, then 
$F_{\pm}^{-1}(\chi_{{}_{Z}}F_{\pm}f)$ cannot even be a measure with finite 
variation. 
\item Let now $\eta$ and $f$ be pdfs, and suppose that 
$F_{\pm}^{-1}(\chi_{{}_{Z}}F_{\pm}f)=0$. 
Then our convergence result  has the following meaning in probability theory:
the series converges in the sense that the expectation values 
of all rapidly decreasing test functions on $X$ are restored. Namely, for 
any rapidly decreasing test function $\psi$ on $X$ we have that:
\[\underset{n\rightarrow\infty}{\mathrm{lim}}\;\int\left(\sum_{n=0}^{N}(I-A_{\eta})^{n}(A_{\eta}f)\right)(x)\cdot\psi(x)\;\mathrm{d}x=\int f(x)\cdot\psi(x)\;\mathrm{d}x.\]
\end{enumerate}
\end{Rem}

It can be easily observed that the condition of our previous theorem 
is not always satisfied for a pdf $\eta$. E.g.\ if $\eta$ is a Gaussian 
pdf centered to zero, then 
it is satisfied, but e.g.\ if $\eta$ is a uniform pdf on a 
rectangular domain centered to zero, then 
the condition is not satisfied. Therefore, one could think that the 
applicability of our deconvolution theorem is rather limited. 
This is not the case, however, as stated in our following theorem.

\begin{Thm}
Let $\eta$ be a pdf on $X$. Then for any $f\in L^{1}(X)$ the series 
\[N\mapsto\sum_{n=0}^{N}(I-A_{P\eta}A_{\eta})^{n}A_{P\eta}(A_{\eta}f)\]
is convergent in the space of tempered distributions, and 
\[\sum_{n=0}^{\infty}(I-A_{P\eta}A_{\eta})^{n}A_{P\eta}(A_{\eta}f)=
f-F_{\pm}^{-1}(\chi_{{}_{Z}}F_{\pm}f),\]
where $Z:=\{y\in X^{*}|F_{\pm}\eta(y)=0\}$. Here $P$ is the parity operator 
on $L^{1}(X)$, namely $Pf(x):=f(-x)$  for all $f\in L^{1}(X)$ and $x\in X$.
\end{Thm}
\begin{Prf}
Let us observe, that if $F_{\pm}\eta$ is real valued and nonnegative for a 
pdf $\eta$, then $\vert 1-F_{\pm}\eta\vert < 1$ is automatically satisfied 
outside $Z$. This is because 
\begin{enumerate}
 \item by our assumption $0 < F_{\pm}\eta$ outside $Z$, thus we conclude 
that $1-F_{\pm}\eta<1$ outside $Z$, and 
 \item by the inequality $\vert F_{\pm}\eta\vert\leq \Vert \eta\Vert=1$, 
we conclude that $0\leq 1-\vert F_{\pm}\eta \vert = 1-F_{\pm}\eta$.
\end{enumerate}

It is easy to see   
that $F_{\pm}P\eta=\overline{F_{\pm}\eta}$ (where the bar 
denotes complex conjugation) for a pdf $\eta$, because $\eta$ is real valued. 
Thus, we have that $F_{\pm}(P\eta\star \eta)=\vert F_{\pm}\eta \vert^2$
is real valued and nonnegative, consequently, by our previous observation, 
the inequality $\vert 1-F_{\pm}(P\eta\star\eta)\vert<1$ holds outside $Z$, 
i.e.\ our previous theorem can be applied by replacing the convolution 
operator $A_\eta$ with the double convolution operator $A_{P\eta}A_{\eta}$. 
\end{Prf}

When applying this theorem in practice, one should take into account 
that the measured 
pdf (which is obtained by histograming in general) is not in the range of the 
convolution operator, but it can be viewed as the sum of a pdf in the 
range of the convolution operator (if our model is accurate enough) and 
a noise term. By the above theorem, the series expansion will be 
convergent on the 
pdf in the range of the convolution operator, but will be divergent 
(most probably) 
on the noise term, as it is not in the range of the convolution operator 
(in general). 
Thus, the problem is that when to stop the series expansion: one should 
let the series 
go far enough to restore the original (unknown) pdf, but should stop 
the series expansion early 
enough to prevent the divergence arising from the noise term. 
This truncation procedure can be viewed 
as a very elegant way to do the high frequency regulation. 
Note, however, that the regulation problem at the finite frequencies 
(at the zeros of the Fourier transform of the convolver pdf) does not 
arise at all, with this method.

The only remaining question is: at which index should one stop to 
keep the noise content lower than a given threshold.

When working in practice, our density functions are discrete in general 
(e.g.\ histograms), thus we may view them as a vector of random variables 
(e.g.\ in the case of histograming, these random variables are the number of 
entries in the histogram bins). Let us denote it by $v$. If A is a linear 
operator (i.e.\ a matrix here), then we have that $\mathrm{E}(Av)=
A \mathrm{E}(v)$ and 
$\mathrm{Covar}(Av)=A\mathrm{Covar}(v)A^{+}$, where we denote expectation 
value by $\mathrm{E}(\cdot)$, covariance matrix by $\mathrm{Covar}(\cdot)$, 
and the adjoint matrix by $(\cdot)^{+}$. Thus, in the $N$-th step of the
series expansion, we have 
\[\mathrm{Covar}\left(\sum_{n=0}^{N}\left(I-A_{\eta}\right)^{n}v\right)=
\left(\sum_{n=0}^{N}\left(I-A_{\eta}\right)^{n}\right)
\mathrm{Covar}(v)\left(\sum_{n=0}^{N}\left(I-A_{\eta}\right)^{n}\right)^{+}.\]
This means that if we have an initial estimate for the covariance matrix 
$\mathrm{Covar}(v)$, we can calculate the covariance matrix at each step, thus 
can calculate the propagated errors at each order.

When using the method of histograming, as the entries in the histogram 
bins are known to obey independent Poisson distributions, 
the initial undistorted estimates $\mathrm{E}(v_{i})\approx N_{i}$ 
($i\in\{1,\dots,M\}$) and $\mathrm{Covar}(v)\approx\mathrm{diag}
(N_{1},\dots,N_{M})$ will 
be valid, where we consider our histogram to be a mapping
$H:\{1,\dots,M\}\rightarrow \mathbb{N}_{0},i\mapsto N_{i}$. 
The squared standard deviations are the diagonal elements of the 
covariance matrix, thus we can have an estimate on the $L^{1}$-norm of the 
noise term at each $N$-th order by taking 
$\frac{1}{\sum_{j=1}^{M}N_{j}}\sum_{i=1}^{M}\sqrt{\mathrm{Covar}_{ii}
\left(\sum_{n=0}^{N}\left(I-A_{\eta}\right)^{n}v\right)}$. 
By stopping the series expansion when this noise content 
exceeds a certain predefined threshold, we get the desired truncation of 
the series expansion.

\begin{Rem}
We show an other (iterative) form of our 
series expansion which may be more intuitive for physicists. Namely, take 
the initial conditions 
\[f_{0}:=A_{P\eta}H,\]
\[\hat{C}_{0}:=A_{P\eta}\mathrm{diag}(H),\quad 
C_{0}:=\left(A_{P\eta}\hat{C}_{0}^{+}\right)^{+}.\]
Then, perform the iteration steps
\[f_{N+1}:=f_{N}+f_{0}-A_{P\eta}A_{\eta}f_{N},\]
\[\hat{C}_{N+1}:=\hat{C}_{N}+\hat{C}_{0}-A_{P\eta}A_{\eta}\hat{C}_{N},
\quad C_{N+1}:=\left(\hat{C}_{N}^{+}+\hat{C}_{0}^{+}-
A_{P\eta}A_{\eta}\hat{C}_{N}^{+}\right)^{+}.\]
Here $H$ means the initial (measured) histogram, $f_{N}$ is the deconvolved 
histogram at the $N$-th step, and $A_{P\eta}A_{\eta}$ is the discrete 
version of the double convolution operator. 
The quantity $\hat{C}_{N}$ is a supplementary quantity, and $C_{N}$ is 
the covariance matrix at each step. The noise content can be written as 
$\frac{1}{\sum_{j=1}^{M}N_{j}}\sum_{i=1}^{M}\sqrt{\left(C_{N}\right)_{ii}}$, 
which should be kept under a certain predefined threshold.
\end{Rem}

\begin{Rem}
As pointed out in the previous remark, one can exactly follow the error 
propagation during the iteration. However, to store and to process the 
whole covariance matrix can cost a lot of memory and CPU-time. Therefore, 
one may rely on a slightly more pessimistic but less costly approximation 
of the error propagation, namely on the Gaussian 
error propagation. This means, that at each step one assumes the 
covariance matrix to be approximately diagonal, i.e.\ this method is based 
on the neglection of correlation of entries (which, indeed, 
holds initially), that slightly will overestimate the error content. 
Gaussian error propagation means that when calculating the action of the 
operators in questions, we apply the following two rules:
\begin{enumerate}
\item if $v$ is a random variable (histogram entry), and $a$ is a number, 
then $\sigma(a\cdot v):=\vert a\vert\cdot\sigma(v)$ (this is exact, of course), 
and \item if $v_1$ and $v_2$ are random variables (histogram entries), 
then $\sigma^{2}(v_1+v_2):=\sigma^{2}(v_1)+\sigma^{2}(v_2)$ (which is 
exact only if $v_1$ and $v_2$ are uncorrelated). 
Here $\sigma$ means standard deviation.
\end{enumerate}
\end{Rem}

\begin{Rem}
Even if the convergence condition for the deconvolution by series expansion is 
satisfied for $A_{\eta}$, it is better to use the double deconvolution 
procedure by $A_{P\eta}A_{\eta}$, for the following reason. In practice the 
measured pdf corresponds to a pdf in the range of $A_{\eta}$ plus a noise term. 
When convolving the measured 
pdf by $P\eta$ before the iteration, the noise level is reduced by orders of 
magnitudes (the convolution by $P\eta$ smooths out the statistical 
fluctuations). As a thumb rule, one iteration step is lost with the convolution 
by $P\eta$, but several iteration steps are gained, as we start the iteration 
from a much lower noise level.
\end{Rem}

%% file: unfold_03.tex
\section{The general case of unfolding}

For the case of general unfolding problems, a series expansion will 
become even more interesting, as there are no known alternative methods 
like the naive deconvolution in the case of deconvolution problems.

Unfortunately, for the general case of unfolding, we cannot state such a strong 
result as for the case of deconvolution. This is because our theorem on the 
deconvolution 
strongly relies on the relation of convolutions and Fourier transformation. 
However, we can state a sufficient condition for the convergence of a 
series expansion for the general case of unfolding. To state this theorem, 
we have to perform studies 
not only on pdfs, but also on probability measures. The spaces $X$ and $Y$ 
are going to denote finite dimensional vector spaces again.

A complex measure $P$ on $X$ is a complex valued $\sigma$-additive
set function defined on on the Borel $\sigma$-algebra of $X$.
The \underline{variation} of the complex measure $P$ is
the nonnegative measure $\vert P\vert$ defined as follows: if $E$ is a 
Borel set, then $\vert P\vert(E)$ is the supremum of 
$\sum_{k=1}^{n}\vert P(E_{k})\vert$ for all splitting $(E_{1},\dots,E_{n})$ of 
$E$, i.e.\ for all such $(E_{1},\dots,E_{n})$ finite system of disjoint 
Borel sets whose union totals up to $E$
(\cite{Matolcsi}, \cite{Dinculeanu}).
The measures with finite variation (i.e.\ the complex measures $P$
for which $\vert P\vert(X)<\infty$) form a Banach space with the 
norm being the value of the variation on $X$, i.e.\ 
$\Vert P\Vert:=\vert P\vert(X)$. Let us denote this space by $M(X)$. 

Recall that a probability measure $P$ on a $X$ is a nonnegative measure 
on the Borel $\sigma$-algebra of $X$, with $P(X)=1$. Thus, a probability
measure is evidently in $M(X)$.

\begin{Def}
We shall call a mapping 
$Q:X\rightarrow M(Y),x\mapsto Q(\cdot|x)$
a \underline{folding measure} if for every $x\in X$ the measure 
$Q(\cdot|x)$ is a 
probability measure on $Y$, and for every Borel set $E$ in $Y$ the function 
$x\mapsto Q(E|x)$ is measurable. 
\end{Def}

Note, that $Q$ may be viewed as a conditional 
probability measure on the product space $Y\times X$.
Evidently, if $\rho$ is a cpdf, then $Q_{\rho}(E|x):=
\int_{E}\rho(y|x)\mathrm{d}y$ defines a folding measure.

\begin{Def}
Let $Q$ be a folding measure $Q$. Then the linear map 
\[A_{Q}:M(X)\rightarrow M(Y),\qquad P\mapsto
\left(\int Q(\cdot|x)\;\mathrm{d}P(x)\right),\]
will be called the \underline{folding operator by $Q$}.
\end{Def}

\begin{Rem} The following remarks are trivial.
\begin{enumerate}
 \item Such an operator is well defined, as for all points $x\in X$ and Borel sets $E$ the 
inequality $Q(E|x)\leq 1$ holds, thus the function $x\mapsto Q(E|x)$ is integrable by any measure with 
finite variation.
 \item By the monotonicity of integration, such an operator is continuous 
and $\Vert A_{Q}\Vert=1$, just as in the $L^{1}$ case.
 \item The folding operator defined above can be viewed as a 
generalization of the folding operator $A_{\rho}:L^{1}(X)\rightarrow L^{1}(Y)$ 
defined by a cpdf $\rho$. This is because $L^1(X)$ can naturally be embedded 
into $M(X)$ by assigning to each $f\in L^{1}(X)$ the measure $E\mapsto P_{f}(E):=\int_{E}f(x)\;\mathrm{d}x$. 
Of course, if the folding measure $Q_{\rho}$ is defined by a cpdf $\rho$, then 
the restriction of $A_{Q_{\rho}}$ to $L^{1}(X)$ is just $A_{\rho}$ as defined before.
\end{enumerate}
\end{Rem}

First, we generalize our deconvolution results to the space of 
measures with finite variation.

\begin{Rem} The convolution of two measures $F,G\in M(X)$ can be defined by 
\[F\star G:\;E\mapsto\int F(E-x)\;\mathrm{d}G(x),\]
where $E$ runs over all the Borel sets. (Of course, $P_f\star P_g=P_{f\star g}$ 
for any $f,g\in L^{1}(X)$.)

The Fourier 
transformations can also be defined on $M(X)$, and have the same properties 
as in the $L^1$ case, except that the Riemann-Lebesgue lemma does not hold 
(i.e.\ the Fourier transform of a measure is a bounded continuous function
but does not tend to zero at the infinity). Therefore, our previous results 
on the series expansion  for the deconvolution (\emph{Theorem 8}) can directly be generalized 
to the probability measures, as the elements of $M(X)$ can also be viewed as 
tempered distributions.
\end{Rem}

As we remarked above for the deconvolution case, we have a powerful result 
also in the more general framework of  measures with finite variation. However,
we are still lacking an answer for the general cases of unfolding.

\begin{Rem} The conditions of the original Neumann series expansion theorem 
fail also in the case of measures.
\begin{enumerate}                                                               
 \item We can observe that our series expansion is only meaningful for 
       the case of a folding operator only when the spaces $X$ and $Y$ are 
the same. (Just as in the $L^{1}$ case.)
 \item Let us assume that $Y=X$. Then, it is easily obtained that a 
       folding operator $A_{Q}$ does not 
       satisfy the required condition $\Vert I-A_{Q} \Vert< 1$, in general. It is 
       trivial by the triangle inequality of norms that 
       $\Vert I-A_{Q} \Vert\leq 2$. We will show now that this inequality can be saturated 
       for a wide class of folding measures. 
       Let $K_{n}(y)$ ($n\in\mathbb{N}$) be a sequence of compact sets with nonzero Lebesgue measure, 
       such that $K_{n+1}(y)\subset K_{n}(y)$ for each $n\in\mathbb{N}$ and 
       $\underset{n\in\mathbb{N}}{\cap}K_{n}(y)=\{y\}$. Let us denote the complement 
       of a set $K_{n}(y)$ by $K_{n}^{\complement}(y)$. 
       Clearly, by considering the splitting $(K_{n}(y),K_{n}^{\complement}(y))$ of the 
       Borel set $X$, one has:
       \[\bigl\vert(I-A_{Q})\delta_{y}\bigr\vert(X)\]
       \[\geq\Bigl\vert\delta_{y}(K_{n}(y))-Q(K_{n}(y)|y)\Bigr\vert+\left\vert\delta_{y}(K_{n}^{\complement}(y))-Q(K_{n}^{\complement}(y)|y)\right\vert\]
       \[=\bigl\vert1-Q(K_{n}(y)|y)\bigr\vert+Q(K_{n}^{\complement}(y)|y).\]
       At the equality, $\delta_{y}(K_{n}(y))=1$ and $\delta_{y}(K_{n}^{\complement}(y))=0$ was used. 
       Let us take the limit $n\rightarrow\infty$ on the right side. 
       By the monotone continuity of measures, we have that $\underset{n\rightarrow\infty}{\mathrm{lim}}\;Q(K_{n}(y)|y)=Q(\{y\}|y)$ and 
       $\underset{n\rightarrow\infty}{\mathrm{lim}}\;Q(K_{n}^{\complement}(y)|y)=Q(X\setminus\{y\}|y)$, furthermore by the subtractivity of measures we have 
       $Q(X\setminus\{y\}|y)=Q(X|y)-Q(\{y\}|y)$. As $Q(\cdot|y)$ is a probability measure, we also 
       have $Q(X|y)=1$. Thus, 
       \[\bigl\vert(I-A_{Q})\delta_{y}\bigr\vert(X)\geq \bigl\vert1-Q(\{y\}|y)\bigr\vert+(1-Q(\{y\}|y)).\]
       As the measure $Q(\cdot|y)$ cannot take up larger 
       values then $1$ on any Borel set, we conclude that 
       \[\left\Vert I-A_{Q}\right\Vert\geq 2\cdot(1-Q(\{y\}|y)).\]
       Thus, if there exists such a point $y\in X$, where $Q(\{y\}|y)=0$, then 
       $\left\Vert I-A_{Q}\right\Vert=2$. 
       When the folding measure $Q_{\rho}$ is defined by a cpdf $\rho$, then 
       $Q_{\rho}(\{y\}|y)=0$ always holds (this is because a measure of the 
       form $P_f$ -- for any function $f\in L^{1}(X)$ -- cannot have sharp points, 
       i.e.\ such points where $P_{f}(\{y\})\neq0$). Thus, 
       $\left\Vert I-A_{Q_{\rho}}\right\Vert=2$ holds for any cpdf $\rho$, 
       therefore the Neumann 
       series cannot converge for $A_{Q_{\rho}}$ in the $M(X)$ operator norm. 
       (But of course, even $Q(\{y\}|y)\leq\frac{1}{2}$ is enough to violate $\Vert I-A_{Q}\Vert< 1$.)
\end{enumerate}
\end{Rem}

Just like in the convolution case, our strategy will be to require much weaker 
notions of convergence. By intuition, one would think that if for all 
$x\in X$ the Dirac-measures $\delta_{x}$ are restored by the method 
(in some sense of convergence), then this would be 
enough for the restoration of any other arbitrary measures with finite 
variation. We provide a similar result with slightly stronger conditions. 
The theorem below is a trivial consequence of Lebesgue's theorem of 
dominated convergence.

\begin{Thm}
Let $A_{Q}$ be a folding operator for some folding measure $Q$. 
Let us fix a Borel set $E$ in $X$. If for all $x\in X$ the sequence 
\[N\mapsto \left((I-A_{Q})^{N+1}\delta_{x}\right)(E)\]
converges to zero, furthermore 
\[\underset{N\in \mathbb{N}}{\mathrm{sup}}\;
\underset{x\in X}{\mathrm{sup}}\;
\left\vert\left(\left(I-A_{Q}\right)^{N+1}\delta_{x}\right)(E)
\right\vert<\infty\]
holds, then for any $P\in M(X)$ the series 
\[N\mapsto\left(\sum_{n=0}^{N}(I-A_{Q})^{n}A_{Q}P\right)(E)\]
is convergent and 
\[\left(\sum_{n=0}^{\infty}(I-A_{Q})^{n}A_{Q}P\right)(E)=P(E).\]
\end{Thm}
\begin{Prf}
First, we note that for any index $N$ the measurable function 
$x\mapsto\left\vert\left((I-A_{Q})^{N+1}\delta_{x}\right)(E)\right\vert$ 
can be bounded by $2^{N+1}$, thus these functions are integrable by any 
measure with finite variation.

We know that for all $x\in X$ the relation 
$\underset{N\rightarrow\infty}{\mathrm{lim}}\;
\left(\left(I-A_{Q}\right)^{N+1}\delta_{x}\right)(E)=0$
holds, furthermore  $\underset{N\in \mathbb{N}}{\mathrm{sup}}\;
\underset{x\in X}{\mathrm{sup}}\;
\left\vert\left(\left(I-A_{Q}\right)^{N+1}\delta_{x}\right)(E)
\right\vert<\infty$.
The integral 
$\int\left((I-A_{Q})^{N+1}\delta_{x}\right)(E)\;\mathrm{d}P(x)$ 
exists for all $N$ and the integrands converge  pointwise to zero as
$N$ tends to infinity. 
As the integrands are dominated by a constant independent of $N$
which is clearly $\vert P\vert$-integrable, by Lebesgue's theorem of 
dominated convergence, the limit and the integration can be interchanged: 
$\underset{N\rightarrow\infty}{\mathrm{lim}}\;
\int\left((I-A_{Q})^{N+1}\delta_{x}\right)(E)\;\mathrm{d}P(x)=
\int\underset{N\rightarrow\infty}{\mathrm{lim}}\;
\left((I-A_{Q})^{N+1}\delta_{x}\right)(E)\;\mathrm{d}P(x)=0$. 
On the left-hand side, $(I-A_{Q})$ can be interchanged with the integration, 
because $I$ is the identity operator and because $A_{Q}$ itself is an integral: 
we can interchange the integrals by Fubini's theorem, namely 
$\int\left(A_{Q}^{N}\delta_{x}\right)(E)\;\mathrm{d}P(x)=\int\int\dots\int Q(E|y_{N})\;\mathrm{d}Q(y_{N}|y_{N-1})\dots\;\mathrm{d}Q(y_{1}|x)\;\mathrm{d}P(x)=\left(A_{Q}^{N}P\right)(E)$, 
for arbitrary power $N$.
Thus, 
$\underset{N\rightarrow\infty}{\mathrm{lim}}\;\left((I-A_{Q})^{N+1}P\right)(E)=0$.

Using the equality $\left(P-\sum_{n=0}^{N}(I-A_{Q})^{n}A_{Q}P\right)(E)=
\left((I-A_{Q})^{N+1}P\right)(E)$, we get the desired result.
\end{Prf}
\begin{Rem}
Assume that the condition of our theorem holds.
\begin{enumerate}
 \item The condition 
       $\underset{N\in \mathbb{N}}{\mathrm{sup}}\;
\underset{x\in X}{\mathrm{sup}}\;\left\vert\left(\left(I-A_{Q}\right)^{N+1}
\delta_{x}\right)(E)\right\vert<\infty$
       (i.e.\ the condition of boundedness) is crucial 
       for the proof in order to be able to interchange the limit and the 
integration. In other words: 
       the restoration of the Dirac-measures $\delta_{x}$ for all $x\in X$ 
is not enough.
 \item If $P$ is a probability 
       measure, then the meaning of our convergence result 
is that the probability of the event (Borel set) $E$ is restored: 
       \[\underset{N\rightarrow\infty}{\mathrm{lim}}\;
\left(\sum_{n=0}^{N}(I-A_{Q})^{n}A_{Q}P\right)(E)=P(E).\]
\end{enumerate}
\end{Rem}

The present theorem is weaker than the one for deconvolution,
nevertheless it provides a computer-testable condition of convergence for any 
unfolding problem (which may not be expressed as convolution). In the next 
section, we shall provide some physical examples which show the method in 
operation. 
Of course, the iteration procedure goes just the same as discussed at the end 
of the previous section.

\begin{Rem}
If we are testing the convergence criterion by computer, some measure theory 
trivialities are useful. Namely, if the condition holds for disjoint sets, then 
it also holds for the union of them. Thus, in practice (e.g. when handling 
histograms), it is enough to confirm the condition when the Borel sets $E$ 
are the histogram bins, because 
then the condition will automatically hold for any set built up from the 
histogram bins. 
Of course, we cannot go below the granulation of our histogram binning, but 
if our granulation is fine enough, the numerical test of convergence condition 
can give an accurate answer.
\end{Rem}

The disadvantage of our presented convergence criterion is that it is rather 
expensive even for a simple 1-dimensional case (however, for a given folding 
measure $Q$, this condition has to be shown only once). It may be better to 
only show the convergence for the \emph{given} unfolding problem, i.e.\ 
on a case-by-case basis, and not for 
the general case of every $P\in M(X)$. (The disadvantage of such a convergence 
condition is that surely it will be violated after a certain iteration step, 
because of the divergence arising from the noise term.) 
Such a condition of convergence may be obtained by Cauchy's root criterion:

\begin{Thm}
Let $A_{Q}$ be a folding operator for some folding measure $Q$. Let us fix 
a measure $P\in M(X)$ and a Borel set $E$ in $X$. If the inequality 
\[\underset{N}{\mathrm{limsup}}\;\sqrt[N]{\left\vert 
\left((I-A_{Q})^{N}A_{Q}P\right)(E)\right\vert}<1\]
holds, then the series 
\[N\mapsto\left(\sum_{n=0}^{N}(I-A_{Q})^{n}A_{Q}P\right)(E)\]
is absolute convergent.
\end{Thm}

With the above condition one may control the convergence of the series 
iteration for a given measured pdf: the condition 
$\underset{N}{\mathrm{limsup}}\;\underset{E}{\mathrm{sup}}\;
\sqrt[N]{\left\vert \left((I-A_{Q})^{N}A_{Q}P\right)(E)\right\vert}<1$ 
may be required as a condition of convergence, where the Borel sets $E$ 
are the histogram bins. Given the order $N$, we shall call the number 
$\underset{E}{\mathrm{sup}}\;\sqrt[N]{\left\vert 
\left((I-A_{Q})^{N}A_{Q}P\right)(E)\right\vert}$ 
the Cauchy index.

\begin{Rem}
The iteration scheme is the same as discussed at the end of the previous 
section (\emph{Remark 9}). In the iteration scheme, the convolution operator $A_{P\eta}$ 
should be replaced by some folding operator $A_{G}$ (used to artificially smear 
the measured histogram in order to reduce the noise content, as pointed out in 
\emph{Remark 11} -- typically this may be chosen 
to be a convolution operator by a Gauss pdf centered to zero, 
or can be chosen to be the identity operator, if smoothing is not needed), 
and the convolution operator $A_{\eta}$ should be 
replaced by the folding operator $A_{Q}$ (describing the physical smearing process).
\end{Rem}

%% file: unfold_04.tex
\section{Examples and applications in physics}

Our first test example will be a deconvolution problem of an initial 
Cauchy pdf of the form $x\mapsto \frac{1}{\pi}\cdot\frac{1}{\Gamma^2+x^2}$, and 
with a Gauss convolver pdf of the form $x\mapsto \frac{1}
{\sqrt{2\pi\cdot\sigma^2}}\cdot \exp\left(-\frac{x^2}{2\cdot\sigma^2}\right)$ 
over the real numbers. We will choose $\Gamma=1$ and $\sigma=1$ in our example. 
By \emph{Theorem 8} we can assure the convergence of the problem. The result 
is shown in \emph{Figure \ref{gausscauchy}}.

Our second test example will be a deconvolution problem of an initial 
Cauchy pdf as in the previous example with a triangle convolver pdf of the form 
$x\mapsto \frac{1}{W^2}\cdot\chi_{{}_{[-W,W]}}(x)\cdot 
\bigl\vert W-\vert x\vert\bigr\vert$ 
over the real numbers. We will choose $W=2$ in our example. 
By \emph{Theorem 8} we can also assure the convergence of the problem. 
The result is shown in \emph{Figure \ref{trianglecauchy}}.

\begin{figure}[!ht]
\begin{center}
{\footnotesize\blopeps[width=15cm, height=10cm]{fig/deconv_gausscauchy/deconv.74.beps}}
\end{center}
\caption{\footnotesize{A Gauss$\star$Cauchy deconvolution by series expansion}}
\label{gausscauchy}
\begin{center}
{\footnotesize\blopeps[width=15cm, height=10cm]{fig/deconv_trianglecauchy/deconv.63.beps}}
\end{center}
\caption{\footnotesize{A triangle$\star$Cauchy deconvolution by series expansion}}
\label{trianglecauchy}
\end{figure}

A signal smearing, caused by a measurement apparatus, is described by folding 
in general. 
In this case the cpdf in the folding integral is the response function of the 
device. Our series unfolding can be applied to remove the non-ideal 
detector smearing at the spectrum level. This is a common issue in analysis of 
recorded data in experimental physics, which may be solved by our method.

Our physical example will be the $\pi^0$ decay. $\pi^0$-s are produced in 
high-energy particle collisions (e.g.\ in hadron or heavy-ion collisions).
The particle $\pi^{0}$ decays through the 
channel $\pi^{0}\rightarrow\gamma+\gamma$ decay ($98.798\%$ branching ratio). 
It has such a short 
lifetime ($8.4\cdot 10^{-17}\sec$), that even in the highest energy colliders 
it only travels at most micrometers before decay, thus from the detector's 
point of view, the resulting $\gamma$ photons come from the collision point. 
The $\pi0$ particles 
are detected via the resulting $\gamma$ photon pairs. This is possible because 
the dominant part of the $\gamma$ yield comes from $\pi^0$ decays in hadron 
or heavy-ion collisions. The $\gamma$ candidate signals are paired to each 
other in every possible combination, and the mass of each pair is calculated 
from the hypothesis that they originate from a common $\pi^0$ decay. The 
combinatorial background is estimated 
by so called event mixing techniques (by taking $\gamma$ candidates from 
different events, thus 
these signals are completely independent). The $\pi^0$ yield as a function of 
momentum thus can be obtained, which plays an important role in high-energy 
particle physics.

However, in certain cases (e.g. in heavy-ion collisions) the reconstruction 
efficiency of $\pi^0$-s can be very low at certain momentum space regions, 
thus this straightforward reconstruction method is not always applicable for 
measuring the momentum distribution of the produced $\pi^0$-s.

A possible idea is to measure the single $\gamma$ momentum distribution, and 
reconstruct the parent $\pi^0$ momentum distribution from it, somehow. 
The arising of the child $\gamma$ photon momentum pdf from a parent 
$\pi^0$ momentum pdf 
is described by a folding, as will be discussed below. The task is: to unfold 
the original $\pi^0$ momentum pdf from the $\gamma$ momentum pdf. This issue 
was also addressed in \cite{Cahn}, however the answer given by the paper 
was not fully satisfactory. Firstly, the method described in the paper was
very specific to the particular case of $\pi^0\rightarrow\gamma+\gamma$ 
decay (and did not deal with 
the general problem of unfolding). Secondly, two kinematical kind of 
approximations were used which are mathematically ill-defined and 
have an unclear physical meaning. It seems, indeed, that our method 
gives a more realistic answer, as it will be shown.

Let us denote the momentum space by $(\mathbb{M}, g)$, where $\mathbb{M}$ 
is a 4-dimensional real vector space, and 
$g:\;\mathbb{M}\times\mathbb{M}\rightarrow\mathbb{R}$ is a Lorentz form
(with signature\ $1,-1,-1,-1$). Let us choose a time orientation on it. 
Let $V^{+}(0)$ denote the positive null cone (positive light cone), 
and let $V^{+}(m)$ be the positive mass shell with mass value $m$ 
($m$ will now play the role of $\pi^0$ mass). The $\pi^0$ momentum pdf is 
defined over $V^{+}(m)$, and 
the $\gamma$ photon momentum pdf is defined over $V^{+}(0)$. However, 
they also can be be viewed as probability measures over $\mathbb{M}$, 
with their support in 
$V^{+}(m)$ and $V^{+}(0)$, respectively. Given a $\pi^{0}$ momentum, 
the $\gamma$ momenta directions (decay axes) are uniformly distributed in 
the $\pi^0$ rest frame (this is the 
physical information put in). Namely, let us take the set 
\[F:=\Biggl\{(p,k)\in\mathbb{M}\times\mathbb{M}\;
\Biggm\vert\;p\in V^{+}(m),\;k\in V^{+}(0),\;g
\left(\frac{1}{\sqrt{g(p,p)}}p,k\right)=\frac{m}{2}\Biggr\},\]
and let us define for every $p\in \mathbb{M}$ the set 
$F_{p}:=\left\{k\in\mathbb{M}\;\mid\; (p,k)\in F\right\}$. Clearly, 
$F_{p}$ is the set of possible $\gamma$ photon momenta arising from a 
$\pi^0$ with momentum $p$ (in other words: $F_{p}$ is defined by the vectors 
in $V^{+}(0)$ which 
have energy $\frac{m}{2}$ in the rest frame of the $\pi^{0}$ with momentum $p$). 
We shall define our folding measure by: $Q(\cdot\vert p)$ is the measure 
over $\mathbb{M}$ for each $p$ which describes the uniform distribution on 
$F_{p}$ (as $F_{p}$ is compact, it has finite measure, thus this is meaningful).
If $P$ is a probability measure over $\mathbb{M}$ describing the $\pi^0$ 
momentum distribution, then the $\gamma$ photon momentum distribution is 
defined by the probability measure 
$A_{Q}P$. Thus, one may try to obtain the parent $\pi^0$ momentum distribution 
by unfolding the measured $\gamma$ momentum distribution. This will be done 
explicitly below for a toy example.

Let us parameterize the momentum space with respect to an Einstein synchronized 
frame $(e_{0},e_{1},e_{2},e_{3})$ that corresponds to the center-of-mass system 
of the collision. We choose the collision axis (the beam axis) 
to be the third spatial coordinate axis which we also call the longitudinal 
direction. 
As the experimental setups of collisions are axially symmetric with respect to 
this axis, the single particle momentum distributions are axially 
symmetric with respect to the longitudinal direction. Therefore, it is 
convenient to parameterize a $\pi^0$ momentum $p\in V^{+}(m)$ in the form 
$\left(g(e_{3},p),\;\sqrt{g(e_{1},p)^2+g(e_{2},p)^2},\;
\arctan\left(\frac{g(e_{2},p)}{g(e_{1},p)}\right)\right)$, and 
a $\gamma$ momentum $k\in V^{+}(0)$ in the form 
$\left(g(e_{3},k),\;\sqrt{g(e_{1},k)^2+g(e_{2},k)^2},\;
\arctan\left(\frac{g(e_{2},k)}{g(e_{1},k)}\right)\right)$. The 
three coordinates are called longitudinal momentum, transverse momentum
and azimuth, respectively. The axial symmetry means that the 
pdfs describing $\pi^0$ and 
$\gamma$ momentum distributions only depend on the longitudinal and 
transverse momentum.

It is even more convenient to introduce a more sophisticated parameterization: 
if $p_{{}_L}$ is the longitudinal momentum and $p_{{}_{T}}$ is the transverse momentum, 
then 
$y:=\mathrm{asinh}\,\left(\frac{p_{{}_{L}}}{\sqrt{m^2+p_{{}_{T}}^2}}\right)$ 
(longitudinal rapidity) and 
$E_{{}_{T}}:=\sqrt{m^2+p_{{}_{T}}^2}$ (transverse energy) can be introduced. 
The so called longitudinal pseudorapidity 
$\eta:=\mathrm{asinh}\,\left(\frac{p_{{}_{L}}}{p_{{}_{T}}}\right)$ is also useful for 
longitudinal parameterization. We 
shall present the pdfs in the $(\eta,\;p_{{}_{T}})$ parameterization.

For demonstration, we take a realistic toy example of $\pi^0$ momentum pdf. The 
$\pi^0$ momentum pdf is characterized by: the momentum pdf of the $\pi^0$ with 
respect to the 
Lorentz invariant measure of the mass shell $V^{+}(m)$ corresponds to 
a product of a Gaussian one in $y$ and an exponential one in $E_{{}_T}$ 
(a typical experimental 
spectrum can be qualitatively described in this a way). The standard deviation of 
the $y$ distribution was taken to be $0.5$, and the inverse slope 
parameter of the 
$E_{{}_T}$ distribution was taken to be $0.5\mathrm{GeV}$.

\begin{figure}[!ht]
\begin{center}
\resizebox{14cm}{!}{\rotatebox{270}{\includegraphics{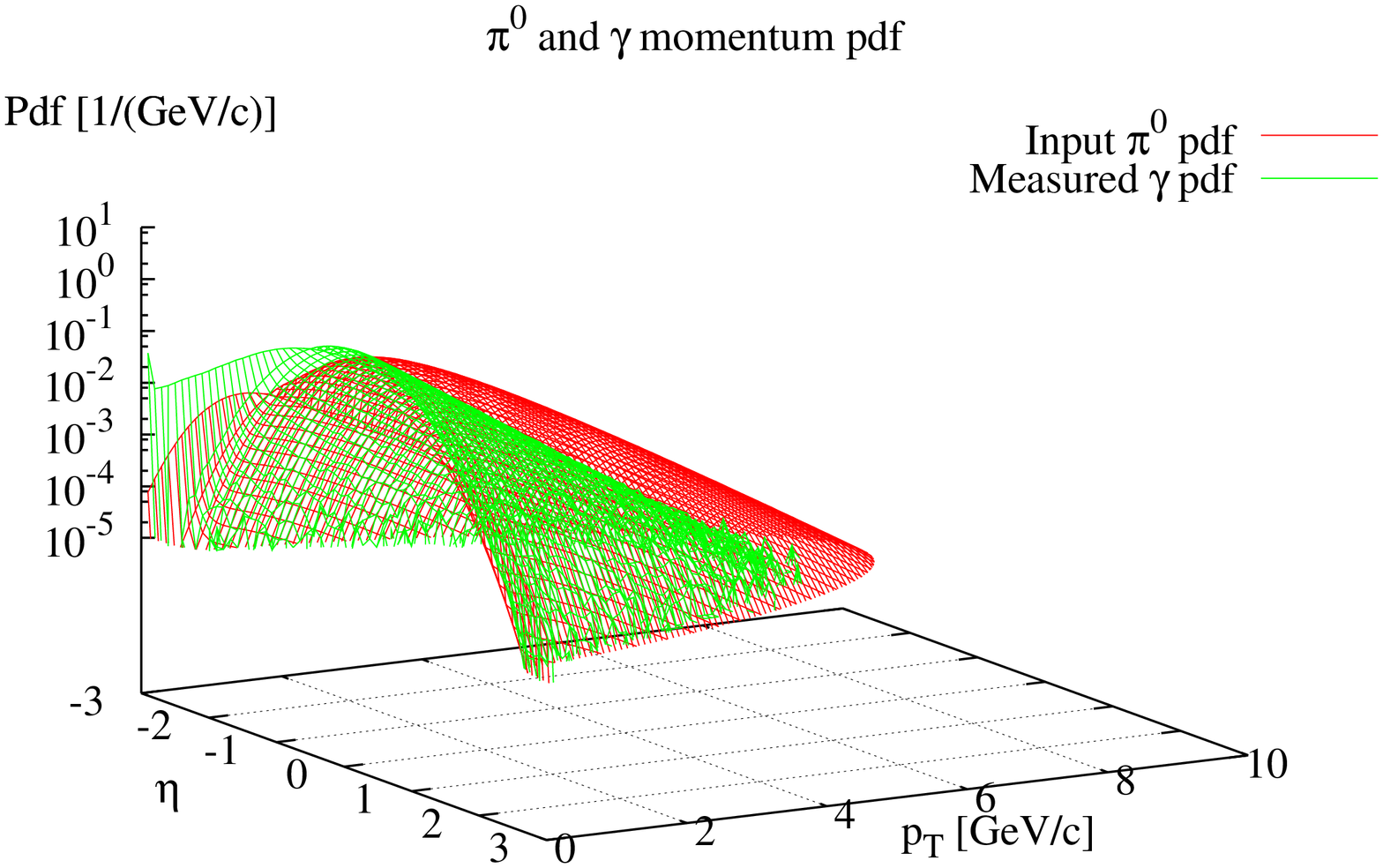}}}
\end{center}
\caption{\footnotesize{Input $\pi^0$ momentum pdf and measured $\gamma$ momentum pdf}}
\label{pi0gamma}
\begin{center}
\resizebox{14cm}{!}{\rotatebox{270}{\includegraphics{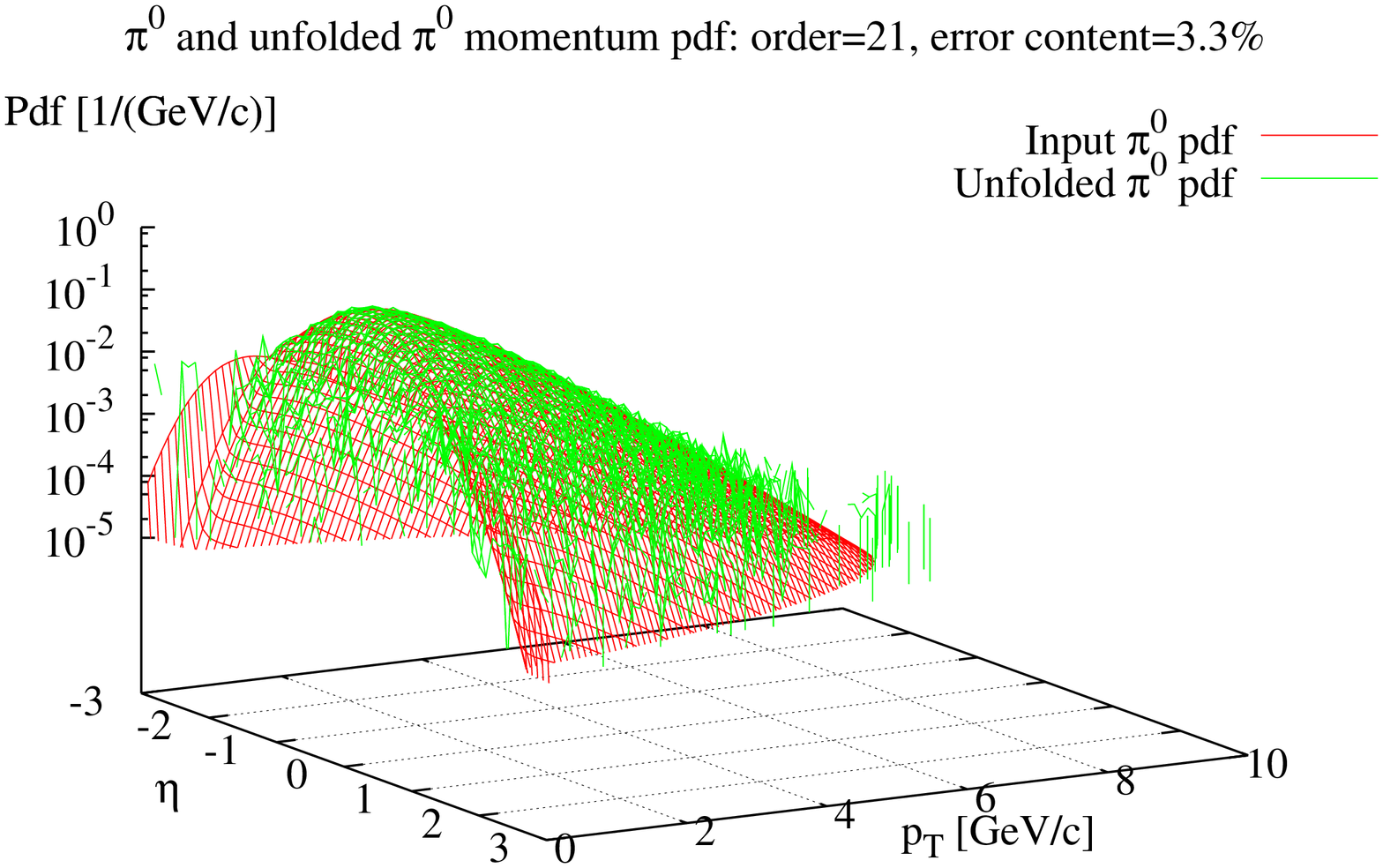}}}
\end{center}
\caption{\footnotesize{Input $\pi^0$ momentum pdf and unfolded $\pi^0$ momentum pdf}}
\label{pi0pi0rec}
\end{figure}

The initial $\pi^0$ momentum pdf is presented in 
\emph{Figure \ref{pi0gamma}} together 
with the arising $\gamma$ momentum pdf. We used a sample 
of $10000000$ Monte Carlo 
$\pi^0$ particles to generate the measured $\gamma$ spectrum.

The unfolded $\pi^0$ momentum pdf is presented in \emph{Figure \ref{pi0pi0rec}} together 
with the initial $\pi^0$ momentum pdf. Due to the high statistics, we did not apply 
smearing for noise reduction (as discussed in \emph{Remark 21}).

To demonstrate the capability of the method, we also included a 
smearing according to 
the CMS-ECAL detector's known energy and angular resolution function, 
when generating the measured gamma responses: the method also removes this 
detector effect from the momentum pdf. This fact is rather important in 
practice, because a non-ideal detector resolution changes the inverse 
slope parameter of the transverse momentum spectrum remarkably, which 
is used in heavy-ion physics to determine the temperature of the 
collided system.

Some sections of the previous pdfs are also presented at 
$\eta=\text{constant}$ slices 
in \emph{Figure \ref{pi0pi0recs0}} and in \emph{Figure \ref{pi0pi0recs1}}.

\begin{figure}[!ht]
\begin{center}
{\footnotesize\blopeps[width=15cm, height=10cm]{fig/unfold_pi0/pi0_pi0rec_s0.21.beps}}
\end{center}
\caption{\footnotesize{Input $\pi^0$ momentum pdf, measured $\gamma$ momentum pdf, and reconstructed $\pi^0$ momentum pdf: taken at the $\eta=0.0$ slice}}
\label{pi0pi0recs0}
\begin{center}
{\footnotesize\blopeps[width=15cm, height=10cm]{fig/unfold_pi0/pi0_pi0rec_s1.21.beps}}
\end{center}
\caption{\footnotesize{Input $\pi^0$ momentum pdf, measured $\gamma$ momentum pdf, and reconstructed $\pi^0$ momentum pdf: taken at the $\eta=0.4$ slice}}
\label{pi0pi0recs1}
\end{figure}

For completeness, we also show the answer given by R. Cahn's prescription (as 
described in \cite{Cahn}), in \emph{Figure \ref{pi0pi0reccahns0}} and 
\emph{Figure \ref{pi0pi0reccahns1}}. Of course, here we did not include additional 
detector effects as in our unfolding case, as R. Cahn's method was not designed 
to undo detector effects. 
As one can see, the reconstructed 
$\pi^0$ momentum pdf given by R. Cahn's prescription is rather far from the 
initial one, especially when compared to the answer given by our series 
expansion method, introduced in this paper.

\begin{figure}[!ht]
\begin{center}
{\footnotesize\blopeps[width=15cm, height=10cm]{fig/unfold_pi0/pi0_gamma_s0.beps}}
\end{center}
\caption{\footnotesize{Input $\pi^0$ momentum pdf, measured $\gamma$ momentum pdf, and reconstructed $\pi^0$ momentum pdf with R. Cahn's method: taken at the $\eta=0.0$ slice}}
\label{pi0pi0reccahns0}
\begin{center}
{\footnotesize\blopeps[width=15cm, height=10cm]{fig/unfold_pi0/pi0_gamma_s1.beps}}
\end{center}
\caption{\footnotesize{Input $\pi^0$ momentum pdf, measured $\gamma$ momentum pdf, and reconstructed $\pi^0$ momentum pdf with R. Cahn's method: taken at the $\eta=0.4$ slice}}
\label{pi0pi0reccahns1}
\end{figure}

\begin{figure}[!ht]
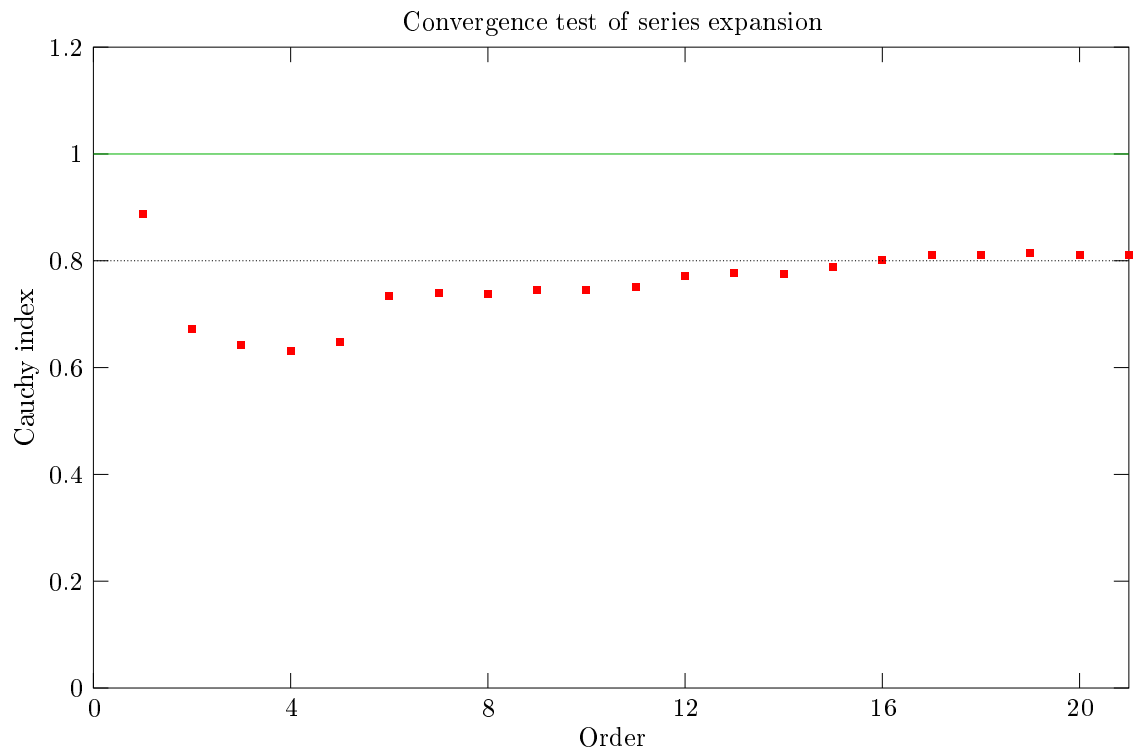

\begin{center}
{\footnotesize\blopeps[width=15cm, height=10cm]{fig/unfold_pi0/cauchytest.beps}}
\end{center}
\caption{\footnotesize{Cauchy convergence test of the series expansion for 
the $\pi^0\rightarrow\gamma+\gamma$ problem}}
\label{cauchytest}
\end{figure}

Our remaining issue is to show the convergence of our series expansion 
for this $\pi^0\rightarrow\gamma+\gamma$ decay unfolding problem. 
In \emph{Figure \ref{cauchytest}} 
we plotted the Cauchy index as a function of the iteration order. It is 
clearly seen that the Cauchy indices are saturating to $\approx0.8$, thus the 
convergence is a consequence of \emph{Theorem 20}.

\begin{Rem}
It is very important to note that when implementing the folding operator, one 
does not have to know the analytic form of the integral. In the 
$\pi^0\rightarrow\gamma+\gamma$ case it 
is possible to calculate the integral formula analytically from kinematics, 
however, the integral becomes very ugly in the $(\eta,\;p_{{}_{T}})$ parameterization. 
Therefore we calculated the action of the folding 
operator by Monte Carlo simulation, which makes the method easy to implement.
\end{Rem}

%% file: unfold_05_ref.tex
\section{Concluding remarks}

A robust iterative deconvolution and unfolding method was developed for 
applications in signal processing. The method has three main advantages:
\begin{enumerate}
 \item It solves any deconvolution problem optimally.
 \item It also solves a wide class of more general unfolding problems 
(for which no general unbiased method was known previously).
 \item The method is quite easy to implement even for sophisticated 
folding problems, if Monte Carlo integration method is applied.
\end{enumerate}

\section*{Acknowledgements}

First, I would like to thank Tamás Matolcsi: clearly, without his inlighting 
lectures and his self-contained books 
I would not even come to the idea of such a solution for the unfolding problem, 
not even mentioning the proof of the results presented here. 
I also would like to thank him 
for valuable discussions, for reading the versions of the manuscript innumerable 
times, and for suggesting several corrections.

I would like to thank to Dezső Varga for the discussion of the problem form 
the physical point of view: 
especially for drawing my attention to the high frequency cutoff role of 
the truncation of the series expansion at finite order.

I would like to thank to Ferenc Siklér for discussions about $\pi^0$ 
momentum spectrum reconstruction in 
the CMS experiment, which lead to the idea of the indirect reconstruction 
from $\gamma$ momentum spectrum: this problem triggered originally the 
surveys presented in this paper, which 
originally were intended to be technical surveys for this particular 
experimental physics application.

I would also like to thank to Bálint Tóth and Árpád Lukács for interesting 
and useful discussions about functional analysis, and to András Pál for 
discussions about the differences of the discrete Fourier transformation to 
the continuum Fourier 
transformation, and the problems which can arise from these in practice.

This work was supported by the Hungarian Scientific Research Fund 
(OTKA, T048898).